# CMOS compatibility of semiconductor spin qubits

## Authors and affiliations

Nard Dumoulin Stuyck[1,2,*], Andre Saraiva[1,2,*], Will Gilbert[1,2], Jesus Cifuentes Pardo[1,2], Ruoyu Li[3], Chris Escott[1,2], Kristiaan De Greve[3,4], Sorin Voinigescu[5], David Reilly[6,7] & Andrew S. Dzurak[1,2,*]

1. School of Electrical Engineering and Telecommunications, UNSW, Sydney, NSW 2052, Australia
2. Diraq, Sydney, NSW, Australia
3. IMEC, Leuven, Belgium,
4. KU Leuven, Leuven, Belgium
5. Edward S. Rogers Sr. Dept. of Electrical and Computer Engineering, University of Toronto, Canada
6. ARC Centre of Excellence for Engineered Quantum Systems, School of Physics, The University of Sydney, Sydney, New South Wales, Australia.
7. Microsoft Quantum Sydney, The University of Sydney, Sydney, New South Wales, Australia
* Corresponding authors: nard@diraq.com, andre@diraq.com, andrew@diraq.com

## Abstract

Several domains of society will be disrupted once millions of high-quality qubits can be brought together to perform fault-tolerant quantum computing (FTQC). All quantum computing hardware available today is many orders of magnitude removed from the requirements for FTQC. The intimidating challenges associated with integrating such complex systems have already been addressed by the semiconductor industry – hence many qubit makers have retrofitted their technology to be CMOS-compatible. This compatibility, however, can have varying degrees ranging from the mere ability to fabricate qubits using a silicon wafer as a substrate, all the way to the co-integration of qubits with high-yield, low-power advanced electronics to control these qubits. Extrapolating the evolution of quantum processors to future systems, semiconductor spin qubits have unique advantages in this respect, making them one of the most serious contenders for large-scale FTQC. In this review, we focus on the overlap between state-of-the-art semiconductor spin qubit systems and CMOS industry Very Large-Scale Integration (VLSI) principles. We identify the main differences in spin qubit operation, material, and system requirements compared to well-established CMOS industry practices. As key players in the field are looking to collaborate with CMOS industry partners, this review serves to accelerate R&D towards the industrial scale production of FTQC processors.

# Contents

# 1. Introduction

## 1.1. The economics of quantum computers

The upfront costs of building quantum computers need to be mitigated for the technology to be economically viable in the long term. The most valuable form of quantum computers will employ fault-tolerant quantum computing (FTQC) and these are estimated to create economic value ranging between 620 billion to 1,270 billion US dollars[1]. However, most of this value is created for customers, with the quantum hardware industry and the quantum software providers taking 10% each of the total market. Since these machines are predicted to entail millions of qubits[2], an economically practical pathway for the development of quantum computers is key.

Besides the upfront costs, the price per quantum computation will have a large impact on the size of its total addressable market. A recent estimate of the energy consumption for running a single instance of Shor's algorithm – one of the most efficient algorithms a quantum computer can run – sets superconducting quantum computers at 38GJ for cracking RSA2048 in 1.5 hours, which is comparable to 20 full car tanks of gasoline[3]. While Shor's algorithm provides useful answers in a single run, quantum chemistry applications, for example, may require hundreds or even thousands of calculations before yielding relevant results. The finite range of applications that could justify such an investment reinforces the need to develop FTQC systems at practical price points.

## 1.2. A quantum era for the CMOS industry

While daunting, these economic challenges are not being faced by engineers and scientists for the first time. Chips with billions of components operating with exceedingly high yield and uniformity are routinely manufactured by the semiconductor industry. This is the result of several decades of exponential growth in computational power at a constant production cost, obtained by miniaturizing components in every new generation, in what is dubbed the Moore's law of transistors[4].

Naturally, the know-how developed by the semiconductor industry would be greatly beneficial to address the economic challenges of building a quantum computer. The upfront costs can be mitigated if the tools and techniques invented and optimized for classical processors can be reused for qubits. Moreover, as exemplified above, quantum algorithms of commercial interest will require multiple runs in quantum computers, each lasting between hours and weeks. For such solutions to be commercially viable it will be necessary to run multiple calculations in parallel, demanding affordable manufacturing of reliable quantum processors on a large scale.

## 1.3. CMOS compatibility of qubits

Most qubit technologies have therefore been retrofitted to be fabricated with standard CMOS tools wherever possible. Photon-based qubits were adapted from free space optical tables to silicon waveguides[5]; superconducting qubits had their substrate changed from sapphire to silicon[6] and ion traps that were originally constructed with laser arrangements are now formed on the surface of grids of gate electrodes on micromachined silicon substrates where radiofrequency pulses are applied to hold ions in place[7].

These adaptations have had varying degrees of impact on qubit quality[8]. Moreover, the exact extent to which each of these systems are compatible with CMOS processes is ambiguous. While fabricating a qubit with a subset of the foundry processes is possible, full integration with multiple metal layers and materials required for dense integration between qubits and with control electronics is out of reach for almost all qubit technologies.

## 1.4. Semiconductor spin qubits in CMOS

While the same questions hang over spin qubits, their degree of compatibility with CMOS processes is naturally higher. Spin qubits are highly coherent in silicon[9] can be fabricated using a gate electrode structure very similar to the field-effect transistors in classical electronics[10] and tolerate many of the manufacturing steps required for a full CMOS process[11]. Curiously, the miniaturization that pushed Moore's law to the current stage has reached the point where quantum effects are inevitable and circumventing them becomes increasingly harder[12]. This is precisely the size limit where individual electrons or holes can be isolated and controlled for quantum information processing.

It is tempting to assume that CMOS technology can readily produce scalable high-quality semiconductor spin qubits. However, two major issues are often overlooked and need to be addressed by both the scientific and engineering communities. Firstly, because of their quantum mechanical nature, semiconductor spin qubits rely on distinctive specifications, such as operating at cryogenic temperatures and extremely limited noise tolerance. Those specifications require strictly defined capabilities and processes – capabilities that are not required for standard, established CMOS technology. Whether those technologies can be adjusted to satisfy the more stringent requirements of quantum bits is the subject of many current studies[11,13–15]. Secondly, the holistic approach that has enabled the scaling of today's CMOS technology, which includes modelling, design, fabrication, and verification, must be revised with qubits in mind. Such an overall design approach is largely missing for FTQC with semiconductor spin qubits, although the first steps towards that goal have recently been made[16,17].

This Review is organized as follows: In Section 2, we introduce the FTQC stack; in Section 3 we give a brief overview of the leading semiconductor spin qubit candidates; in Section 4 we discuss spin qubit device compatibility with CMOS industry; in Section 5 we highlight the challenges to scale to a spin qubit based FTQC; and finally, in Section 6 we summarize the main conclusions.

# 2. Fault tolerant quantum computing stack with spin qubits

Like the classical computer engineering stack, quantum computers will rely on a stack of technologies with varying degrees of abstraction, see Fig. 1. The stack layers range from the qubit hardware plane all the way to the user application layer. Unlike digital computers, however, the structure and performance of the full stack is reliant on the same essential element – the ability to correct errors.

## 2.1. Requirements of fault tolerance

Quantum computations rely on analogue operations on qubits, exposing them to errors. The purpose of fault tolerant quantum computing is to identify and control collective properties of qubits that can be effectively treated as digital[18]. However, such error correction comes with a significant overhead, requiring a full layer of quantum error correction (QEC) machinery to suppress errors and ensure reliable operation, see Fig.1 and Box 1. Several QEC protocols have been proposed and are currently being investigated in small-scale prototype quantum systems[19–22].

A minimum fidelity for all the qubit operations is required for error correction to provide a protection against errors that can be systematically improved. This error threshold varies depending on the particularities of the types of errors, qubit arrangements and error correction protocol, but typical target error rates are well under an error in a hundred operations. Most estimates provided in this paper assume an error in a thousand operations. Silicon-based spin qubits have demonstrated operations at this level in multiple platforms[23–27].

The caveat is that even with exquisitely high qubit quality, a large overhead on the number of physical qubits per logical qubit is required, see Box 1. As an active field of research, QEC requirements are steadily improving. Yet, useful universal fault-tolerant quantum computing presently requires qubit counts in the millions[28,29]. Attempts at pushing part of the error correction into the physical qubit layer by creating more noise-resilient qubit hardware are ongoing[30] but still need to be combined with algorithmic error correction, as described above[31].

## 2.2. Quantum computing stack

Similarly to a classical computer, a future quantum computer will require different layers of hardware and software to operate in unison, as depicted in Fig. 1a. At the lowest level, a quantum chip holds the physical implementation of the quantum processor. Qubit options currently being investigated include semiconductor spin qubits, superconducting qubits, photonic circuits, and trapped ions, among others[32,33]. Above the quantum layer, a quantum-classical hardware layer addresses the input/output signals of the qubits. Its role is to generate classical, analogue signals which perform basic qubit operations, such as qubit initialization and control, and measure output analogue signals that correspond to the qubit readout. Furthermore, this layer routes (classical) qubit readout signals to a dedicated QEC layer which uses the information to update the operation parameters for the next cycle or track the errors for post-calculation correction in accordance with the implemented error correction or mitigation protocol. In this way, and at an abstract level, the QEC layer groups physical qubits into fault-tolerant logical qubits with suppressed error rates. A sense of the complexity required for QEC implementations can be gained by looking at the example of the surface code, one of the most well understood schemes[19,20], see BOX 1. Finally, a top layer allows users to program and compile quantum algorithms on logical qubits using dedicated quantum programming languages and compilers.

It is instructive to draw a comparison between this FTQC stack and the design abstraction layers describing advanced computer chips, shown in Fig. 1 b). In a traditional CMOS design process, clients and developers first decide on the high-level system specifications such as an overall system architecture, cost, and performance. Specifications are then translated into a more tailored system architecture with specific modules such as digital, analogue, and mixed signal, power management, memory, and communication modules. Module functionality is constructed using logic (Boolean) gates and translated into transistor circuits. At the very last step, the devices are drawn in a layout compliant with the manufacturer's design rules[34].

This holistic top-down approach to designing complex systems is an iterative process where implications of layout choices, such as speed and power consumption, can feed back into the design of the layers above. The iterative process is only made possible because of the computer tools available, which help to design at any level of abstraction, and to efficiently bridge interfaces between the different levels. The system complexity of FTQC is such that it will necessitate a similar co-optimization approach. Current research into FTQC architecture is performed exclusively in the opposite direction, with a bottom-up structure where the properties of the qubit set the functionalities and requirements for the full system.

## 2.3. Modularity vs integration

The vast majority of the scaled up architectural planning for quantum computers entail the combination of modules that are interconnected by some form of coherent quantum link that allows quantum entangling operations across modules[35–37]. This modular approach alleviates problems with the density of inputs and outputs per modules, increases the yield of quantum modules, and reduces the impact of module-wide catastrophic error events such as impacts from cosmic rays.

On the other hand, the reliance on coherent quantum links is challenging due to the fragility of the architecture to errors introduced in those links. While entanglement distillation protocols can improve the performance of coherent linking, the overhead introduced by such protocols becomes prohibitive at even moderate scale computers[38]. Moreover, the increase in power consumption and size of the equipment poses difficulties for the economic viability of quantum computations, reducing the range of applications for which running a quantum computation is worth the investment[39].

Semiconductor spin qubits offer an alternative architectural pathway where most of the system can be integrated in the same package. This tight integration leverages the small sizes of spin qubits, the advanced integration processes from the CMOS industry and its unprecedented yield and uniformity. Moreover, depending on the fabrication conditions of the qubit platform, parts of the control system can be co-integrated on the same chip as the qubits. Still, challenges are present when trying to adapt this technology for the extreme demands of qubit performance and the complex instruction set required for quantum information processing as detailed in Section 4.

## 2.4. Variability-aware architecture for scalable operation

Fault tolerance requires scaling qubit control for arrays of millions of qubits with strict requirements on operation fidelity, which demands the development of variability-aware architectures. For small quantum processors, each individual qubit can be individually wired and controlled with electromagnetic pulses of arbitrary shapes, catering to the individual properties of every qubit. Forgoing individualised control for a more scalable approach creates an antagonism between the variability of the qubit parameters and the overall fidelity of the operations.

In classical electronics, the general observation known as Rent's rule specifies that the number of control lines $N_L$ shouldn't scale at the same rate as the number of active components, which would in this case refer to the number of qubits $N_Q$[40]. Instead, some power dependence $N_L \propto N_Q^p$ with $p < 1$ is desirable. There are two identifiable approaches to tackle this problem.

<u>First approach: Uniform qubits with shared control</u>

The first approach relies on optimizing spin qubit devices to the highest levels of uniformity, such that all quantum operations can be delivered with bitline-wordline addressing by applying control signals through lines that are shared between multiple dots[41,42]. This approach, inspired by the dense arrays implemented in NAND memory architectures, roughly reduces the scaling of the number of input and output lines to the square root of the number of qubits, $p = 0.5$. Realistically, the full architecture would also require the establishment of interconnected modules of qubits and additional on-chip apparatuses for monitoring the processor performance, enabling bootstrapping and other tasks, leading to some value of $0.5 \leq p \leq 1$.

The main challenge then becomes optimizing the materials and architecture to reduce device variability sufficiently. Variation in this case directly impacts the fidelity of the operations, which exponentially impact the overheads for fault tolerance.

The sources of disorder for spin qubits are related to those for classical CMOS technologies and include charge traps, strain, device defects, and interface roughness[43]. Quantum dots trapped in quantum wells, further away from gate electrodes and dielectric interfaces are typically less susceptible to these sources of disorder than dots formed against the $Si/SiO_2$ interface. The progress in high quality Si/SiGe quantum well interfaces is most notable with demonstrated measurements of statistical variability of 300mm wafer reaching yields for quantum dot formation as high as 99.8%[44].

Despite improvements in *quantum dot* yield, scaling quantum technologies will also require metrics for the still to-be-developed concept of *qubit* yield, with spin physics playing a dominant role. For example, while the spin state is partially protected from most of the disorder present in semiconductor devices, specific materials and control techniques can affect its coherence[45]. For instance, electric field driving of spins under a magnetic field gradient created by a nanoscale magnet leaves spin states more susceptible to charged impurities and charge noise[46]. Magnetic field control methods, on the other hand, do not have this disadvantage but have the drawback of slower qubit operation.

As a high degree of qubit device uniformity is crucial for a shared control approach, significant resources must be invested in characterizing and reducing the levels of disorder and the spin susceptibility to these effects. While semiconductor foundries are taking the lead in terms of materials, gate uniformity and charge noise levels[11,13,47], there is still no exact correlation between the traditional yield of CMOS transistors and the yield of the spin qubits[47].

Second approach: Variable qubits with individual control

The second approach relies on compensating at least partially the qubit variability with some integrated system-on-chip that generates or routes analogue control signals that best fit the individual qubit properties. This approach, combined with the use of robust control techniques to mitigate higher levels of variability and disorder[48,49], can lead to good spin qubit yield but demands qubits to be fabricated in a process that can also yield classical electronics circuits. In the search for more uniform quantum dot systems, some qubit implementations have taken distance from these materials, which sets constraints on their integration with conventional integrated circuits.

Taking an approach of co-integrating qubits with classical electronics may require relatively high qubit operating temperatures of a few Kelvin limited by the system heat density and limited available thermal conductivity. An additional benefit for operating at elevated temperatures is that cryostats have orders of magnitude higher cooling power at a few Kelvin compared to their lowest millikelvin temperatures. The operation of electron and hole spin qubits has already been demonstrated above 1K[50] [26,51,52], albeit with somewhat reduced fidelity compared to qubit operation at milliKelvin. The exact impact of this reduction of fidelity on the performance of error correction needs to be investigated.

Another challenge of this approach is on the co-integration of cryo-electronic circuits that can reliably implement the instruction set for error corrected architectures on the one hand, and the dense qubit array on the other. Individually wiring qubits in a dense array will push the limits of interconnect pitch in the front-end-of-line processes in CMOS foundries, and lead to complications in RF crosstalk between lines.

Robust control

Both architectural principles have challenging target qubit uniformity and would benefit from the development of robust control techniques based on an in-depth analysis of the statistical variability of quantum dots spin qubits. In order to understand the minimum requirements to address and correct unavoidable variations in multiqubit systems[53], a combination of mass characterisation of qubits and sophisticated solid-state modelling tools is required.

The robust control techniques also need to be capable of improving qubit performance without adding further constraints on qubit addressing or control bandwidth. Techniques for global control[48,49], physics-based pulse shaping for error suppression[54] and methods for exploiting the non-Markovianity of noise[55] are some of the examples of such generic methods for evading errors from known physical sources that can be applied equally to all qubits in an array.

## 2.5. Integration with classical electronics

Most semiconductor spin qubit demonstrations so far still make use of room temperature electronics for qubit control, which are connected to the qubit stage through metres of specialty wires. A typical experimental setup includes microwave sources for spin control, low-noise DC voltage sources for gate electrode biasing, fast arbitrary waveform generators for quantum dot potential and barrier control, and digitizers for spin readout. Devices are often individually packaged on custom-made PCBs using wire bonding, and signals are conducted to the devices through cables that must be accurately filtered and thermalised at the various stages within the cryostat.

Regardless of the final architectural principles adopted for a full-scale spin-based quantum computer, the issue of density of I/O lines becomes prominent. An architecture that does not leverage the integration with electronic circuits would require the transfer of low-noise analogue signals from room temperature controllers to the qubit chip. Scaling above a few hundred qubits in this manner would require an impractical amount of wiring, components, and associated heat load for commercial cryostats. To control millions of qubits, on-chip or in-package integration of the classical and quantum layer would form an attractive alternative and may even be required to achieve the necessary wiring density, illustrated in Fig. 2. Semiconductor spin qubits may be able to benefit from various levels of integration offered by the classical computing industry.

As a general principle, integrated classical electronics negatively impacts the qubit performance, either by heating the device, imposing unfavourable fabrication steps, or generating spurious electromagnetic fields that act as noise on the qubits. The level of integration is ideally kept at the minimum demanded by the bandwidth of the interfaces, see Box 2.

Four levels of electronics integration can be envisioned – minor circuits interspersed between qubits, larger modules acting as system-on-chip, integration of separate chips with high-density interconnects as a system-in-package and finally cabled connections between chips arranged in separate temperature stages of a cryostat, including room temperature. It is likely that all four levels of integration will be needed depending on the task for which the components are designed.

The first difficulty with the tighter on-chip integration levels is the harmonisation between the qubit fabrication and the co-integrated electronics. Complex circuitry requires component yields that are essentially 100%, which can only be achieved using relatively aggressive steps such as high temperature anneal and strong doping. For design architectures where classical and qubit circuitry occupy separate modules on chip, masking steps may be able to protect the silicon area where qubits will be fabricated against some of the fabrication steps. However, this becomes more challenging if classical electronic components are interspersed between qubit structures, for example to route signals to qubits in a field-programmable manner.

An alternative to the co-integration of qubit and classical electronics and classical electronics is heterogeneous integration of chips fabricated with separate processes. With advanced semiconductor packaging, it is possible to provide dense interconnects between chips fabricated in different batches, using different silicon technologies and even from different manufacturers. This type of integration benefits from significant economic push from classical computing systems, such as graphics processing units, AI and mobile hardware, which are increasingly leveraging system-in-package approaches.

At the scale of millions of qubits, power dissipation and thermal budget management within a chip or between chips becomes an important consideration. CMOS qubits in general perform better at lower temperatures and even the mild effects of temperature on spin qubits are sufficient to cause concerns over a blow up in physical-per-logical overheads for fault tolerance.

Finally, there is the matter of the overall efficiency of the quantum computer. The total cost of quantum computations will ultimately be determined by its power efficiency, and the integration of electronics at low temperature requires dissipating the heat in an inefficient way. Even if the cryocooler serving the quantum computer operates at the Carnot limit, i.e. at the theoretical maximum efficiency, the cooling power efficiency for a fixed dissipation rate per qubit scales with the ratio between room temperature and the temperature of the stage where the electronics is integrated. Maximizing the number of components at room temperature or at least at higher temperature stages is therefore beneficial for the overall costs per computation.

## 3. Semiconductor spin qubit flavours

In its simplest form, a semiconductor spin qubit is formed by the spin state of a single electron or hole which is confined in every dimension by a potential. The two main approaches to confine charge carriers are donor implantation (*e.g.*, P in bulk Si) producing an atomic-like 3D isotropic confinement or trapping the charge carrier electrically against a dielectric barrier (such as $SiO_2$ or $Si_{1-x}Ge_x$)[56]. This last method is referred to as an electrostatic quantum dot (QD).

The individual charge carriers isolated in these nanostructures can be leveraged for quantum computation in different ways. Their charge states (or position in space) were among the first ideas explored theoretically for qubit encoding[57,58]. However, the electric noise ubiquitous to solid state devices degrades the quantum properties of charges too fast for any meaningful operations. Instead, the spin of the carrier has much better coherence times and can be used as the qubit as long as an external magnetic field is applied to define a quantization axis. A spin pointing parallel or antiparallel to the external field represents a 0 or 1 in this scenario. Other alternatives explore the spins of two or three electrons and how they align relative to each other. These types of qubits are commonly referred to as single spin qubits (or Loss-DiVincenzo qubits), singlet-triplet qubits and exchange-only qubits, respectively. Table 1 gives an overview of the leading spin qubit designs and we refer to Ref.[56] for further information about these spin encodings.

Fig. 3 shows the most important energy scales for semiconductor spin qubits, and their typical range. We focus here on electrons, but similar considerations are due for hole-based qubits. To avoid decoherence and ensure controllable qubit operation, the qubit's spin states need to be sufficiently spaced from other energy states. Higher excited states composed by orbital ($E_O$) or valleys ($E_v$) excitations need to be kept above the Zeeman splitting ($E_Z$). The orbital excitation is associated with the quantum confinement of the carrier, which can be engineered by controlling the nanometre-scale geometry of the device. The valley excitation is more challenging – breaking the cubic symmetry of silicon through strain or confinement only partially lifts the valley degeneracy of the silicon conduction band. The symmetry between the conduction band minima along opposite crystallographic axes can only be broken by a singular potential such as the Coulomb potential of a dopant or the sharp offset of the conduction band introduced by an interface. The valley excitation energy is typically an order of magnitude smaller than the orbital excitation and dominates many of the error processes in qubit control and readout. It is worth noting that standard CMOS devices operate at volt-scale while quantum devices, owing to the mentioned energy scales, operate at a μV-scale.

In addition, the thermal energy ($k_B T$ where $k_B$ is Boltzmann's constant) is restricted to avoid thermal excitations to higher states. This limits semiconductor qubits, so far, to temperatures lower than 4K (or less depending on the technology), well below the typical operating regime of conventional CMOS.

Each semiconductor spin qubit implementation integrates methods to initialize, read out and control the spin states. One-qubit gates are typically driven by ac signals delivered either electrically using a gate electrode close to the quantum dot (in combination with micromagnets or other artificial spin-orbit coupling) and referred to as electric dipole spin resonance (EDSR), or magnetically, by an antenna or dielectric resonator referred to as electron spin resonance (ESR). Alternative spin qubits schemes only requires broadband signals for qubit operation[59]. Two-qubit gates in each spin qubit implementation are enabled by spin-exchange interactions between neighbouring quantum dots[60]. Scalable architectures must be able to switch exchange interactions electrically between ON/OFF regimes by moving the electron wavefunctions 'closer' and 'farther' apart – either in real space or by virtue of additional electrical signals.

The characteristics of the quantum dots and the qubits themselves are strongly linked to material substrates, dielectrics, electron confinement and architecture, see Table 1. Readout and control methods are typically adapted to the specificities of each implementation. An extensive review of the physics of spin qubits and their implementations can be found in Ref.[9] In this work we focus on their compatibility with standard CMOS industry processes and capabilities, and discuss challenges for scalability.

**Table 1:** Overview of leading semiconductor spin qubit candidates for FTQC and key differences.

| Potential formation | Quantum dots with lithographically-defined gates | | | | Donor atoms |
|---|---|---|---|---|---|
| Schematics (adapted from Ref. [56])<br><br>**Legend**<br>🟩 Silicon<br>⬛ Dielectric<br>🟦 Gates<br>🟪 Charge density | 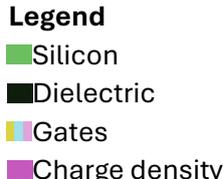 | | 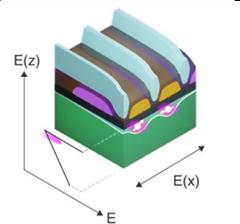 | | 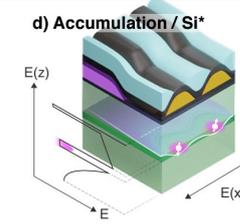 |
| Host material | Si/SiO$_2$ (SiMOS) | | Si/SiGe QW | Ge/SiGe QW | Si |
| Nuclear spin free isotope | $^{28}$Si | | $^{28}$Si, $^{70}$Ge, $^{74}$Ge | | $^{28}$Si |
| Charge carrier | Electron | Hole | Electron | Hole | Electron/Hole |
| Gate-to-gate pitch or half QD pitch (nm) | < 50 | 50 – 100 | 50 – 100 | 100 – 200 | 5 – 10 |
| AC control (B- Magnetic E - Electric) | B, E | E | E | E | B, E |
| Technologies | Planar, FinFET, FD-SOI | Nanowire, FD-SOI | Planar, FinFET | Planar, nanowire, nanohut | Bulk |
| Commercialization | Diraq, Quobly, Quantum Motion, Equal1, SemiQon, Hitachi | Quobly | Intel, Arque | - | Silicon Quantum Computing |

The two main arguments used by the quantum computing community to claim CMOS compatibility as a selling point for semiconductor spin qubits are: 1) the use of semiconductor materials; and 2) the use of lithographically defined metallic gates to confine charge carriers. Both arguments, at least in principle, would lend themselves for fabrication of spin qubits in the advanced, well controlled processing environments used for state-of-the-art CMOS – and thereby benefit from the tremendous investments in process control and stability in such environments. In the next section, we review the status of the quantum device and quantum control layer. We focus on state-of-the-art semiconductor spin qubit devices and highlight their similarities and differences to classical CMOS technologies from this perspective. Next to the quantum device layer itself, FTQC will likely require powerful classical CMOS for QEC and qubit control. Delivering qubit control signals from room temperature electronics is hard to scale in terms of I/O and other connections, hence an attractive proposition would be for this control layer to operate in unison with qubits at cryogenic temperatures, despite this being outside of the typical CMOS operating regime[61,62].

# 4. Current qubit device compatibility with CMOS manufacturing

Both semiconductor spin qubits and CMOS technology are umbrella categories, spanning a wide range of variation in device materials, designs, and applications. In this section we highlight the applicability of current state-of-the-art CMOS technologies for semiconductor spin qubits. We start with discussing the general compatibility and then focus on specifics of the leading semiconductor spin qubits introduced in the previous sections.

## 4.1. CMOS technologies for spin qubit devices

Interestingly, modern classical transistors, with channel dimensions of ~10 nm × 5 nm × 50 nm, are small enough to form quantum dots. Each FinFET and FDSOI MOSFET of minimum gate width behaves as a single-electron transistor (SET), or single-hole transistor (SHT), with pronounced Coulomb peaks that can be observed in the transconductance, $I_{DS}$-$V_{GS}$, at cryogenic temperatures, up to 50 K[63–65]. These dots have large enough energy spacing of a few meV between the first and second orbital energy levels to also host spin qubits. Such QD behaviour has been observed in commercial 3 nm, 5 nm, 7nm[65], 16 nm FinFET[66] and in 22 nm[63,67] and 28 nm FDSOI transistors[64].

Hence, from the perspective of reusing the most advanced CMOS technology, manufacturing qubits using technology derived from commercial FinFET (or nanosheet), and FDSOI technologies would appear attractive because of their feature size and threshold voltage tuneability. Nonetheless, Process Design Kits (PDKs) for these technologies are released with manufacturability restrictions designed to maximise yield of conventional devices which, however, limit the types of qubit arrays and qubit gates that can be readily implemented.

In general, all semiconductor spin qubit flavours to date have material stacks deviating from standard industry designs. To different degrees, they will require substantial effort to integrate into mature technology nodes or make them compatible in terms of fabrication. As of today, the majority of commercially available, classical CMOS and related technology nodes are built on a $HfO_2$/$SiO_2$/Si or $HfO_2$/$SiO_2$/SiGe based interfaces as part of the device and gate stack[68]. Few exceptions exist for niche applications such as high-power, RF and optoelectronics based on materials including SiGe, GaAs, GaN, GaSb, InSb, AlN, SiC, InP[68,69]. Advanced nodes below 28nm currently offered are limited to fully depleted[70–72] (FD) silicon on insulator (SOI), FinFET[73,74] or nanosheet MOSFET[75].

While metal gates and linear arrays of thousands of single QD's are relatively straightforward to manufacture, non-trivial changes are necessary to form coupled QDs and 2D arrays of coupled QDs, as will likely be needed for a universal quantum processor unit (QPU). The most important required changes are:

    (i)    blocking dopant diffusion and contact formation between gates to enable coupled QD formation,

(ii) reducing gate pitch to allow for strong tunnel coupling in exchange gates,
(iii) reducing gate leakage current without increasing short channel effects and the gate voltage beyond 1 V, which would increase the power consumption of the control electronics, and
(iv) isotopic purification of Si and Ge in the QD array channel.

The gate pitch reduction for two-qubit gate control is critical and most invasive in terms of the mask and process changes and is explored in depth in Section 4.5. To implement interstitial gates similar to academic qubit prototype devices[60] in a 3 nm FinFET process one needs to both block doping and contact formation in the source/drain contact regions between gates and reduce the gate pitch. In technologies where the required dot gate pitch is not directly achievable, the restriction can often be mitigated, for example by the use of selective back gate in FDSOI CMOS[63]. Improved manufacturability of lower dimension and lower spacing selective back gates are likely required to optimise the number and location of tunnel coupled qubits as well as their electrical parameters within array architectures.

## 4.2. Si-MOS quantum dots

The well-known Si-MOS stack comes with significantly different requirements for quantum computing purposes compared to traditional CMOS technologies. The transistor gate-stack is optimized for a combination of maximal gate control (transconductance) while limiting direct gate leakage currents – justifying a trend towards high-k dielectrics and FinFET or gate-all-around (GAA) structures such as nanosheets. In contrast, Si-MOS spin qubits might suffer from the higher trap densities of high-k dielectrics[76]. Si-MOS qubits are formed at the Si/SiO$_2$ interface near the gate stack, increasing the impact of the exact gate stack material. Charge noise originating from the gate stack can couple to the spin state through spin-orbit interaction and impact qubit coherence and control operation fidelities[13,77–82]. Furthermore, a coefficient of thermal expansion (CTE) mismatch between the gate material and the interface can result in strain at cryogenic temperatures, which can lead to band deformation and creation of spurious quantum dots[83,84]. Therefore, gate stacks must be further optimised for uniform device strain, either by proper choice of materials, dielectric thickness, or gate geometry or a combination thereof[85].

Commercial CMOS technologies use a substrate of naturally occurring silicon $^{nat}$Si. About 4.7% of $^{nat}$Si consists of $^{29}$Si, an isotope that carries nuclear spin and therefore can contribute to magnetic noise that causes decoherence of the electron spin states[56,86]. While not an issue for classical electronics, the presence of nuclear spinful isotopes limits quantum coherence of spin qubit devices[87]. Switching to isotopically enriched (nearly nuclear spin-free) $^{28}$Si improves electron spin coherence times by an order of magnitude by removing the magnetic coupling to the uncontrolled nuclear spin bath[60,87]. Deposition of $^{28}$Si on a 300mm scale has been demonstrated in different R&D fabs[11,47,88,89], but it is currently not a commercially offered option.

## 4.3. Quantum well-based quantum dots

Besides MOS and MOS-like spin qubits, recent academic progress on germanium quantum well-based hole spin qubits is most impressive, enabled by the growth of high-quality undoped strained Ge quantum wells[90]. However, Ge quantum wells are typically deposited on a Si$_{1-x}$Ge$_x$ strain-relaxed buffers, which are not commonly used in the commercial semiconductor industry. Such 'virtual substrates' can be created from a real Si substrate in different ways, such as depositing a thick (> 1 μm) Ge and Si$_{1-x}$Ge$_x$ (> 1 μm) layer, or using a pure Ge wafer and growing a thick Si$_{1-x}$Ge$_x$ layer (~ 2.5 μm) before quantum well deposition[91,92]. Deposition rates are limited due to the required low growth temperature for high quality interfaces[93,94]. Furthermore, to avoid strain relaxation in the quantum well, processing steps are limited below ~ 500 °C after quantum well growth, bringing additional processing challenges of creating high yielding gate stacks[92,95]. Despite often resurging research interest in Ge as a transport layer in some variants of CMOS, the quality of the Si-oxide interface remains much higher than that of Ge-oxide, making a full technological switch to Ge-based CMOS highly unlikely.

Like Ge quantum well devices, Si/SiGe QW spin qubits require low defect-density pristine interfaces for qubit operation. Namely, the energy splitting of otherwise degenerate valley-states of the electrons, which is crucial for the isolation and operation of spin qubits, depends critically on the quality of this interface[86,96]. The requirement for finite $E_V$ to lift the valley degeneracy goes beyond the more modest requirements for suppression of inter-valley scattering mechanisms considered for mobility increase in classical CMOS technologies – making reuse of those (classical) learnings difficult. Due to strong out-of-plane electrical fields and higher band offsets between Si/SiO$_2$ vs. Si/SiGe, $E_V$ is larger for SiMOS compared to Si/SiGe QW structures. Proposals to increase valley-splitting energies in Si/SiGe QW spin qubit devices rely on a further tuning of the Si/SiGe quantum well heterostructure[78,97,98]. Manufacturing Si/SiGe QW qubits requires a virtual substrate made of a thick graded SiGe layer (> 1 μm) before deposition of a $^{(28)}$Si quantum well[55,99]. The maximum processing temperature after Si quantum well deposition is around 750 °C[92], somewhat higher than for Ge QW devices but below commonly used annealing and activation temperatures used in mainstream processes. SiGe is routinely used in commercially available foundry technologies, but its use, unlike as a quantum well in quantum devices, is rather limited to that of a channel material to improve hole mobility and threshold voltages for p-type devices.

## 4.4. Dopant atoms spin qubits

Spins of electrons bound to a donor dopant atom exhibit some distinct benefits for qubit implementation, including a high valley-orbit splitting, high charging energy, repeatable single spin properties and natural electrostatic stability. Another potential avenue would be to adopt acceptor impurities and their hole spin as qubits[101,102]. Two general approaches for the deterministic fabrication of single- and few-dopant devices are being explored in literature; a top-down approach based on the implantation of individual ions and a bottom-up approach leveraging scanning tunnelling microscope tips to create atomic-scale lithographic patterns that can be used to selectively incorporate the dopant. Both technologies strive to accurately define the dopant position and the number of incorporated dopants in each site.

The ability to control the dopant count and position is a key requirement for spin-based quantum computing. In variants of this architecture that target direct exchange between spins in the dopants as the means for entangling gates, the strong wavefunction oscillations caused by valley interference create significant challenges. In particular, a misplacement of a pair of donors by even a single lattice site is sufficient to create changes in the exchange coupling between the pair of spins by orders of magnitude[103].

Reliable donor two-qubit operation requires either tight fabrication tolerance or more robust two-qubit gate implementations. Strategies to circumvent this problem may include the development of deterministic implantation methods based in scanning tunnelling microscope lithography[104], the adoption of spin-spin interaction methods that do not rely on the contact exchange interaction[105,106] or the combination of using clusters of dopants implanted with a non-deterministic method and calibrating a posteriori the spin qubit properties conditional on the interaction with the donor nuclear spins. For instance, in the weak-exchange limit, a resonant CNOT gate can be performed without requiring a precise value of the exchange interaction[107]. Nuclear two-qubit gates with >99% fidelity, and three-qubit electron-nuclear Greenberger-Horne-Zeilinger (GHZ) states have been obtained by imparting a geometric phase to an electron bound to both nuclei[27], again without requiring a precise placement of the two donors. Alternatively, the dopant confinement can be combined with some gate-based movement of the charge to enable the qubit-qubit entangling gates [56,105,108].

Despite its challenges for scalable manufacturing, significant progress has been made in single- and two-qubit gate operations[27,109–112] and on the demonstration of precise donor placement[113,109,114,115] which encourages the search for fabrication techniques compatible with VLSI processes.

## 4.5. Advanced lithography for spin qubit devices

Crucial for all quantum dot semiconductor spin qubit types is the voltage control of the exchange interaction which enables two-qubit gates[56]. As mentioned before, the highest two-qubit gate fidelity can be achieved through a dedicated barrier gate electrode between two quantum dots which voltage-controls the exchange interaction strength[23,116–119]. Because the exchange interaction strength scales inversely with quantum dot separation, this sets an upper limit on the electrode gate pitches. For reference, in the original Burkard, Loss and DiVincenzo proposal for a coupled Si-MOS QD quantum gate, a QD-to-QD spacing of 30 nm was assumed[120]. When accounting for the barrier gate to be placed between the gate electrodes defining the quantum dots, a yet-to-be-developed technology with a gate pitch of 15 nm and gate length of ~10 nm is required – well beyond current state of the art in commercial lithography.

The necessary gate pitch for sufficient exchange interaction strength strongly depends on the chosen material and technology and are summarized in Table 1 and contrasted against commercially available technologies in Fig.4. For example, in recent demonstrations of fault-tolerance compatible two-qubit gate fidelities for spin qubits in Si-MOS quantum dots, a gate pitch of 30 nm was used[25]. As a comparison, the commercially available 3nm FinFET CMOS technology features a gate pitch of 45 nm[121]. Several strategies to overcome the gate pitch scaling challenge are discussed next.

To tackle this scaled gate pitch challenge, a first method used by academic groups for Si-MOS, Si/SiGe QW, and Ge/SiGe QW prototype spin qubit devices is gate overlapping with multi-layer gate patterning[122–124]. QD and barrier gate layers are patterned sequentially and interlayer isolation is achieved through gate oxidation[124] or oxide deposition processes[125]. Overlapping gate qubit devices have successfully been replicated in custom 300mm process flows combining deep ultraviolet (DUV) and e-beam lithography in Si-MOS[84,126] and FD-SOI[127] platforms. The combination of different lithography methods increases layout flexibility but unfortunately brings additional challenges in pattern alignment, uniformity, and wafer throughput that all require addressing.

Recently, alternative routes have appeared to achieve scaled pitches using processes beyond typical academic fabrication capabilities. Those include a double gate layer patterning with DUV on a FD-SOI platform effectively creating a 40 nm gate pitch[128], DUV on finFET Si-MOS platforms[11,129,130], and single layer patterning on a Si-MOS platform using extreme ultraviolet (EUV) lithography[131] or single year patterning on a Si/SiGe platform using e-beam lithography in combination with a single layer back-end-of-line (BEOL) process[59,132]. As these processes mature and scale an important milestone will be to demonstrate compatibility with high-fidelity single- and two-qubit control as required for QEC.

# 5. Scaling challenges for full-stack CMOS QC

A semantic variation of the term *scaling* is being observed over time. In the original context of the microelectronics industry, the scaling laws of transistors referred to the increase in computational power at a fixed total resource budget. For instance, a total chip area, total manufacturing cost or total power consumption[68].

In modern applications it is common to use the term scaling to refer to an increase in size, regardless of the increase in resources required. Examples of this type of scaling up in classical computing industry include the combination of processors to form a computational cluster or the combination of data centres to form a cloud infrastructure. These forms of integration form bandwidth bottlenecks and are not equivalent to the scaling of compute power within a chip.

An analogy to these distinct forms of scaling up can be drawn in quantum computing architecture.

## 5.1. System requirements

Adapting a holistic VLSI design approach for semiconductor spin qubits requires defining strict system requirements and the ability to efficiently mitigate changes at every design level[34] shown in Fig. 1b. Currently, both pose significant challenges for semiconductor spin qubits.

The use-cases for quantum algorithms highly depend on the size of the quantum hardware. The large number of qubits required for FTQC algorithms inspired researchers to look for practical use cases for so-called noisy intermediate-scale quantum (NISQ) computers in the interim[133,134]. While impressive in principle, demonstrations of quantum supremacy using highly specific NISQ algorithms suffer from a cat and mouse game as matching (or even surpassing) classical solutions are quickly developed[135–137]. Near-term value for NISQ algorithms hasn't been found yet, thus far confirming earlier resource estimates and no-go theorems[135,138,139]. The required algorithmic depth of 'real' applications precludes using uncorrected qubits thus far, exacerbating the need for error correction and FTQC. Although FTQC system requirements can change radically and rapidly as quantum algorithms and QEC resource requirements improve, at present, realistic estimates still indicate a need for millions of physical qubits in order to solve meaningful problems[39].

## 5.2. Classical cryo-CMOS electronics

Since commercially available (classical) CMOS integrated electronics are designed to operate at or near room temperature conditions, cryogenic effects (such as dopant deionization/freeze-out, reduced lattice cooling and hot carrier effects, threshold shifts and so on) could in principle jeopardize cryogenic operation of CMOS electronics.

Luckily, the most advanced production level FinFET and FDSOI CMOS technologies feature undoped strained Si and SiGe channels for *n*- and *p*-MOSFETs, respectively, which do not suffer from dopant deionization at cryogenic temperature and exhibit even significantly improved transistor figures of merit for digital, analogue-mixed-signal, and RF circuit performance at 2 K compared to 300 K[144,145]. Since FDSOI, FinFET and SiGe BiCMOS circuits exhibit better performance with lower power consumption at cryogenic temperatures[145], moving the control electronics from room temperature to cryogenic temperatures is in principle beneficial for all QPUs. However, power density and power consumption remain harsh constraints. Commercially available pulse tube cryostats have cooling powers of 2 − 4 W at 4 K, which would set an upper bound on the power dissipation budget allowed for the control electronics in a QP with 1 million qubits to less than 4 μW/qubit, which is a rather aggressive target for high-fidelity, low-noise electronics. Higher power, custom built cryostats exist and would improve such thermal budget by about three orders of magnitude, however, thermal conduction would still limit the maximum sustainable power density[146].

A number of cryogenic circuits operating at microwave frequencies up to 200 GHz have been demonstrated in recent years in planar bulk CMOS[147,148], 22nm and 14nm FinFET CMOS[149–152] and FDSOI CMOS[63,144,150,153–158], and some even with a few integrated QDs[63,64,144,153,154,158]. However, a cryogenic CMOS solution encompassing all requirements for operating the qubit layer (initialization, control, and readout) is still missing. Furthermore, most demonstrations fall far short by a factor of ~1000 of the power dissipation target for commercially available and spin qubit proven cryogenic setups. A survey of the most recent cryogenic circuits for quantum control and readout can be found in Ref. [157] and Fig. 5 shows an overview of the estimated power consumption per qubit for different cryo-CMOS control demonstrations in different CMOS technologies. It's important to note that co-integration of cryo-CMOS and spin qubits is a young field of research and that the comparison between datapoints in Fig. 5 is not strictly one-to-one. Industry standards have not yet been put into place and the existing cryo-CMOS demonstrations all differ in qubit technology scope (i.e., superconducting, spin qubits, etc), capabilities (i.e., qubit control, readout, etc.), operation speeds, and signal generation flexibility. It is expected that power consumption can be lowered by tailoring the cryo-CMOS specifications to the qubit device requirements with a system technology co-optimization approach, for example by adjusting gate stacks and/or threshold voltages to optimize system level performance[158]. A complication, at present, is the absence of open or freely available cryo-CMOS PDKs, that would allow realisation of designs optimized in terms of power consumption and performance.

### 5.3. Modelling and design of integrated quantum devices

The increasing complexity of QP prototypes creates a need for more advanced software tools for design and modelling, following the example of Very Large-Scale Integration (VLSI) design tools. When integrating qubits with cryogenic control circuits on-chip, the number of integrated devices can jump from below 100 to well over 1,000,000. Classically, this level of complexity is well handled by commercially available custom-IC design and verification tool suites. It must be noted that the required tool sets and expertise to use them comes at significant cost compared to those typically used for fundamental device research, and that classical chip design, even using advanced design tools, requires dedicated teams of highly skilled designers. Such designers do not currently exist and will likely need to be trained, with a combined classical-quantum skillset. An additional expense is incurred as high complexity necessitates that most of the effort be shifted from design to verification to have confidence that fabrication runs will yield and function as expected[159].

Key to the scientific developments thus far supporting semiconductor-based spin qubits is the theoretical modelling behind the interpretation of experimental results. Besides connecting the observed phenomena with their microscopic origin, these developments have led to the creation of tools to describe the solid-state physics of qubit devices in a regime that is significantly different to the environment of typical semiconductor device applications.

The lower temperature of operation in quantum processors creates the first problem. The thermal energy broadening in room temperature devices is leveraged in the context of the Thomas-Fermi model to accelerate the convergence of simulations. The sharp change in the Fermi-Dirac statistical energetic distribution of carriers below 4K poses a challenge to simulate effects near the chemical potential of the carriers.

This model also encounters challenges in the context of bound electron states in nanometric devices, in which the strong electron-electron repulsion creates quantum correlations that are ignored in mean-field theories such as Thomas-Fermi, or even Hartree-Fock and Density Functional Theory. The correct understanding of the device electronic structure often evokes the use of advanced quantum chemistry tools, such as full configuration interaction or quantum Monte Carlo.

Another problem is the impact of the atomic structure of devices on the wavefunction of individual electrons, which can only be described through costly atomistic simulations such as Tight Binding or *ab initio* theories.

One challenge in the use of existing commercial processes and tools is that the simulation models and full process development kits (PDKs) of transistors are typically only calibrated over a temperature range of -40°C to +125°C. To accurately simulate these commercial processes at cryogenic temperatures, custom models need to be developed, requiring advanced cryogenic modelling and measurement expertise.

Modelling of individual transistors or qubit structures can be assisted with tools such as COMSOL Multiphysics or Synopsys TCAD, although is often driven by development of bespoke and multi-physics, across-boundary research tools[85,105,160–166] to assess characteristics such as superconductivity, magnetic fields, oxide quality, and qubit parameters[167,168,168–170]. An active field of research is developing software toolsets that can co-simulate classical electronic circuits together with quantum hardware[85,171,163,172].

## 5.4. Volume testing of quantum metrics

As qubit arrays increase in size, volume testing will become a strict requirement[11,14,15,173]. Firstly, for the most efficient mapping of the quantum algorithm to the physical qubit layer, the performance of each qubit should be mapped in the control hardware layer to account for non-uniformities in qubit operation or coherence performance. Next, for qubit fabrication, the statistical performance of a large volume of qubits is crucial to enable feedback to the manufacturing process to further improve the qubit device quality. Volume testing for semiconductor spin qubit devices is challenging because of their operation at cryogenic temperature ranges and lack of established test protocols. Furthermore, the qubit device structure is also complex as many control electrodes and peripheries, such as charge sensors and the spin controlling ESR antenna or EDSR magnet, are needed for the complete characterization of a single qubit. The characterization of qubit devices can be categorized as direct testing and indirect testing.

For direct testing of full qubit properties, such as coherence times and qubit gate fidelities, volume testing is very laborious[56,174]. Not only do the spin qubits operate at cryogenic temperatures, the typical timescale for fully optimizing a single few-qubit device in research labs can take up to days. The recent transition from lab-based single spin qubit devices to CMOS industry scale fabrication brought the need for statistical analysis of spin qubit metrics [11,14,15,173]. However, the best efforts so far are still manual based individual device preparation and characterization on a handful of qubits. To address this challenge, solutions can be sorted from different directions. On the cryogenic environment part, recent work on qubits beyond one Kelvin provide a pathway to reduce the thermal cycle time by one order of magnitude (from ~ 10hr to mK to ~ 1hr to K) while allowing extraction of full qubit coherence metrics including two qubit gate operation fidelities[26]. On the qubit control part, development of automated device tuning could enable fast and parallel measurements. Additionally, demonstrations of efficient statistical analysis of qubit metrics with limited devices show the device-to-device variations can be effectively captured on limited devices by exploiting different operation conditions[175]. Recent demonstration of on-chip multiplexing and switching matrix routers offer an alternative pathway towards high-volume qubit characterizaiton[176,177].

With above challenges in direct testing, many of the efforts are devoted to indirect testing with faster turnaround to infer the qubit (or part of the qubit structure) performance. Moreover, standard CMOS and more straightforward electronic transport testing strategies can be applied on test structures similar to qubit devices, which can facilitate pinpointing the limiting qubit operation factors or assessing process impact in the qubit fabrication flow. For example, Hall mobility measured using macroscopic Hall bar structures has been used as a metric to benchmark the quality of the qubit housing interface. The peak mobility and percolation density have been used as a guideline for gate stack optimization[13]. Industry standard automatic probe stations have also been applied to qubit devices. At room temperature, the device electrode and ohmic contact yield can readily be verified. Furthermore, cryogenic automatic probe stations operating around 1.6 K have been developed for 300mm wafers[178,179]. They can provide fast learning on some of the key qubit structure performance, allowing rapid learning cycles for fabrication, characterization, and analysis for process optimization. The relative low temperature also enables change sensing down to the last electron in the qubit quantum dots, giving insight into the device uniformity in the qubit operation regime, which can provide valuable guidelines for the design of larger-scale qubit architectures and insights in operation requirements.

Given the nanoscale nature of spin qubit structures, volume testing along with statistical analysis can reveal some of the key device physics, which can otherwise be hidden by local random variations because of a limited number of datapoints. The threshold voltage of the last electron correlates well with the transistor threshold in the many electron regime[180] resembling same nonuniformity source and pointing towards charge impurities. In the milli Kelvin range, the transport-based SET measurements are much faster than qubits, while its charge noise has been shown to well resemble qubit noise. The statistical analysis of SET charge noise in different working conditions and devices shows good agreement with a standard two-level fluctuator model, which could help understanding the origin of the charge noise and provide further optimization directions.

The CMOS fabrication on qubit devices and the volume testing are mutually reinforcing better understanding from the volume testing enables fab device optimization while better devices from the fab with higher uniformity makes the automated device testing easier to implement on a larger volume. Nonetheless, challenges remain: the quantum computing community uses different methods to assess basic qubit metrics such as qubit coherence and gate fidelities, which makes cross-system benchmarking hard[25,181]. The link between qubit metric to the qubit fabrication process is not as straightforward as standard CMOS testing techniques. We believe these challenges can be best addressed with larger datasets of qubit metrics mating with more comprehensive physical models, as we discussed in the previous section.

## 5.5. Development cost of design, development and running a FTQC

As of today, no truly fundamental barrier has been encountered that would prevent the development of a semiconductor spin qubit-based FTQC. However, the main scaling challenge might be overcoming the cost of developing processes that allow to design and fabricate one. Footprint and power consumption considerations suggest that, out of the commercially available nodes, only the most advanced ones are suitable for this task.

Aside the cost of modifying a commercial process, the remaining costs can be subdivided into two parts: the technology development cost, and the cost of design and fabrication in the respective technology.

Technology development cost

The development costs of commercial CMOS nodes, including R&D costs, are well kept industry secrets, but are estimated in the tens of billions of dollars per node[182]. This upfront investment imposes significant challenges for the economic viability of general-purpose foundries despite the half a trillion dollar a year classical semiconductor market[183]. It is, therefore, unlikely that completely new nodes would be developed solely for FTQC applications. However, modest modifications of commercial CMOS might be viable[184]. This is the main reason why the degree of CMOS compatibility is key for the successful fabrication of a quantum processor architecture in commercial foundries.

Cost of production and design

Cost of full designs in available nodes has increased exponentially with technology node and is currently well beyond the budget of typical academic groups or small startups, see Fig. 4. Adapting CMOS flows and optimizing fabrication yield will come with tantalizing costs and long timelines. Introducing radically new tools or materials will amplify this problem even further. The spread of focus due to a lack of dominant qubit device design will make it even harder to push for changes in the traditionally conservative semiconductor industry. Therefore, academic groups and startups are now partnering with industry leaders to co-develop processes and lift some of the financial challenges[17].

In view of the costs of developing novel CMOS technology nodes, and the volumes required for foundries to recover such costs, the per-wafer costs of CMOS have also increased rapidly over the past technology nodes, making a direct, dedicated series of CMOS runs difficult to swallow for individual groups or companies. Nevertheless, such 'per wafer' cost still pales compared to the full development cost – which can be written off over other, non-quantum customers, making the CMOS compatibility of Si spin qubits an attractive aspect in the economic analysis of upscaling.

An additional cost component is the necessary deviation from standard CMOS for FTQC operation – a cost that cannot be written off over non-quantum customers and therefore needs to be taken up fully by the quantum players.

For some (older) technology nodes, and for (planned) pilot line facilities worldwide, an attractive alternative to dedicated wafers is using so-called 'multi-project-wafers', where different customers share the cost of silicon and thereby reduce the production costs. However, combining multi-project wafers with 'dedicated', non-standard technology will require thorough coordination and road mapping between a priori competing quantum bit efforts to tailor the processes in such a way that multi-process wafer operation is feasible at all.

Qubit fabrication in CMOS typically requires breaking some of the manufacturer design rules, which adds costs and reduces their ability to leverage existing software. Therefore, the degree of CMOS compatibility is a key aspect for the economic viability of qubit technologies. Spin qubits stand out as the only technology that is natively based on semiconductors, accepting traditional materials and processes from the industry, and imposing only minimal constraints to the advanced packaging of multiple chips.

# 6. Conclusion

A fault tolerant quantum computer has the potential to revolutionize several areas of our society. However, reaching that potential requires scaling hardware from a few tens to a few million high-quality qubits, based on available quantum error correction codes and quantum algorithms. Manufacturing such high numbers of devices has only been demonstrated in the semiconductor industry. Therefore, leveraging that industry might be the only realistic pathway towards fault tolerant quantum computing. From all possible quantum information processing candidates, semiconductor spin qubits share most similarities with industrial standard processes, materials, and design. However, in this review we highlighted some of the important differences which so far have prevented a direct and full use of the available industrial capabilities.

Interestingly, the development of semiconductor spin qubits might profit from several CMOS industry efforts to maintain Moore's law. Motivation for low power next generation technology partially coincides with requirements for cryo-CMOS electronics. The search for novel transistor materials such as (Si)Ge can bring solutions for the heterogenous co-integration of CMOS-based components and quantum well-based qubits. Development of AI tools to manage chip design complexity can help make advanced nodes more accessible to semiconductor spin qubit startup companies[185]. Advanced BEOL and 2.5D/3D technology, such as backside power delivery, could help solve QPU interconnect and IO bottlenecks, if shown to be compatible with cryogenic operation and spin qubit control.

# 8. Figures

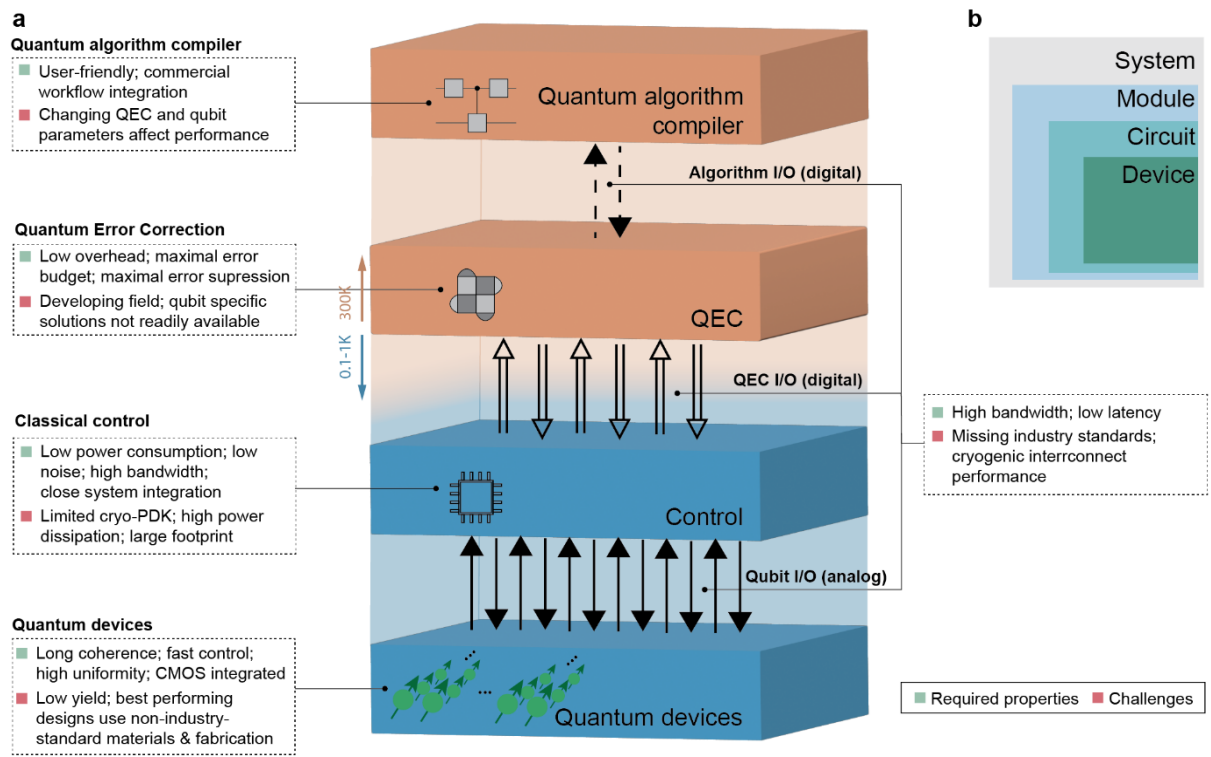

**Fig. 1 Schematic of the different layers inside a FTQC. a**, Fault-tolerant quantum computing stack, indicating the different functional layers. **b**, VLSI design abstraction stack.

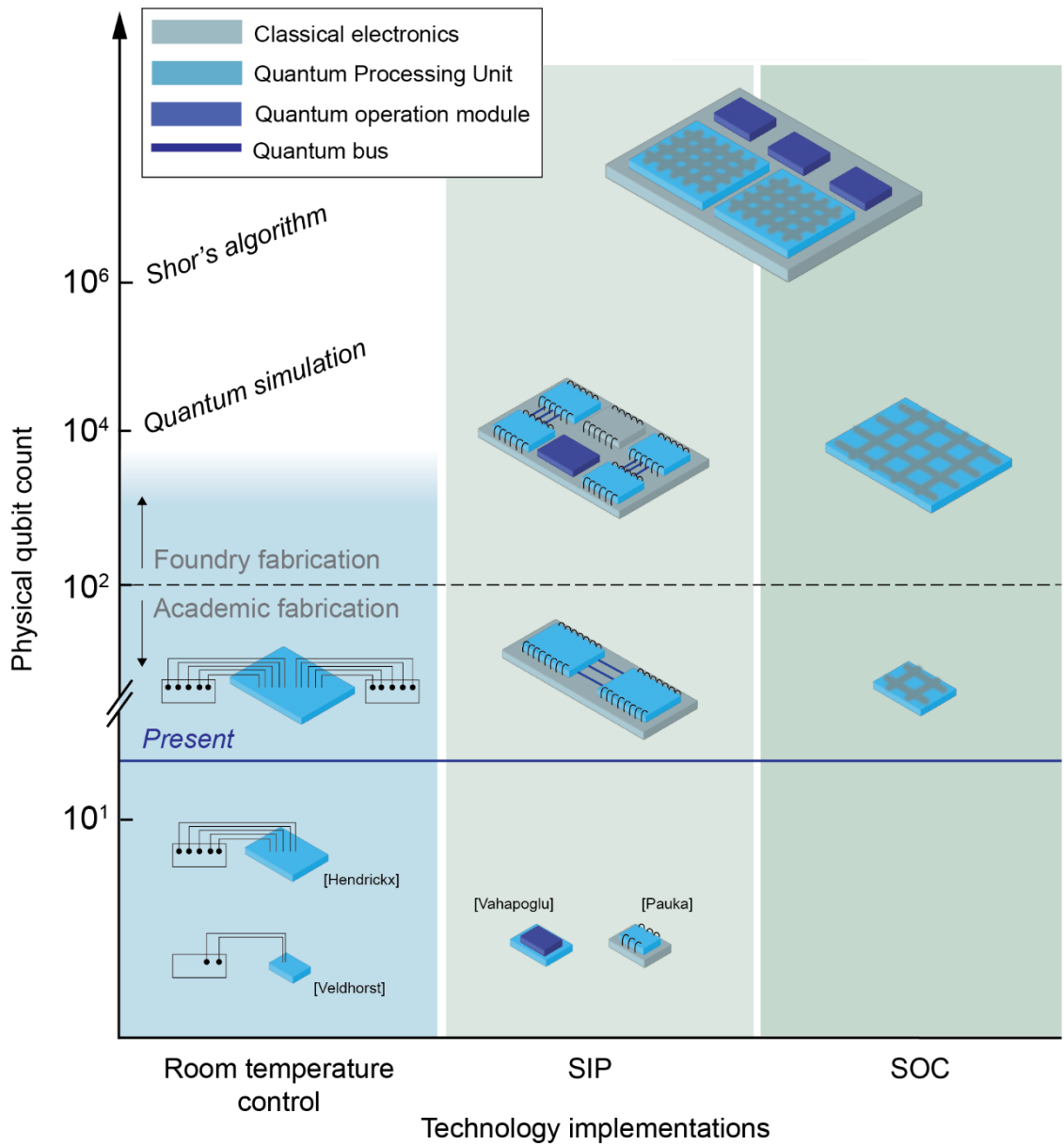

**Fig. 2 Possible semiconductor spin qubit roadmap.** Highlighting the different technology regimes including room temperature control, system-in-package (SIP), and system-on-chip (SOC) implementations, and their estimated achievable qubit counts. Semiconductor spin qubit devices fabricated in academic cleanroom environments are limited in qubit count due to low device yield, indicated with dotted grey line. Estimated physical qubit count to achieve quantum simulation and Shor's algorithm.

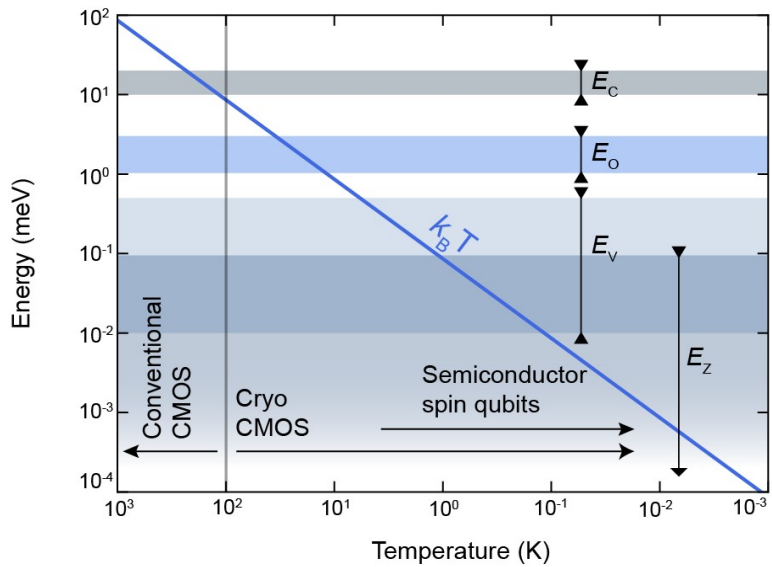

**Fig. 3 Relevant energies and temperature domains for conventional CMOS, Cryo-CMOS, and semiconductor spin qubits.** Shown are typical energy ranges for charging ($E_C$), orbital ($E_O$), valley-splitting ($E_V$), and Zeeman ($E_Z$) energies.

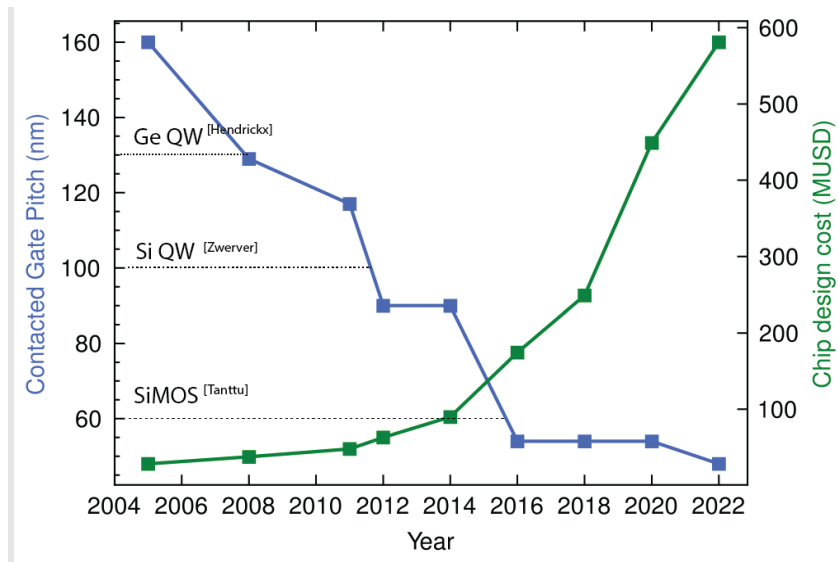

**Fig. 4 Contacted gate pitch in nm and example chip design cost in million USD for commercially available technology nodes.** Dotted lines indicate pitch requirements for different spin qubit technologies[11,25,186]. Data points from Refs. [182,187].

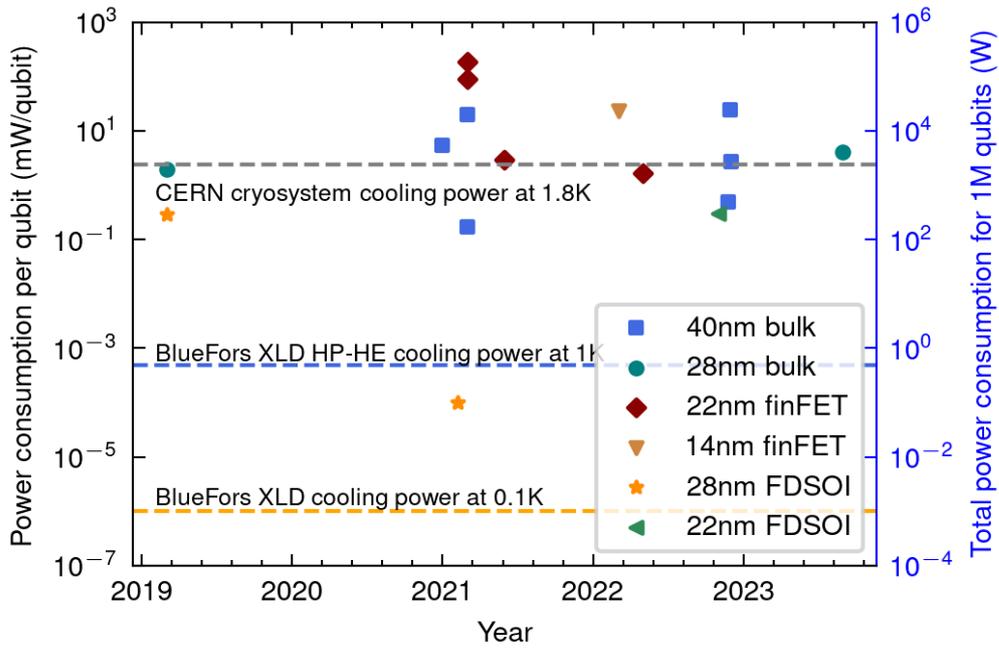

**Fig. 5 Cryo-CMOS control chip power consumption per qubit**. Overview of recent cryo-CMOS qubit control chip demonstration's power consumption normalised per channel, i.e. per qubit. It's important to note that the existing demonstrations differ in scope and specifications, see main text for details. Horizontal dotted lines indicate the maximum cooling powers for state-of-the-art cryo-systems operating at temperatures of 0.1K, 1K, and 1.8K. The BlueFors cryo-systems are commercially available while the CERN system is custom-developed. Data from Refs.[147–152,188–193,158,64,146].

# 9. Boxes

**Box 1**

## Quantum error correction with surface codes

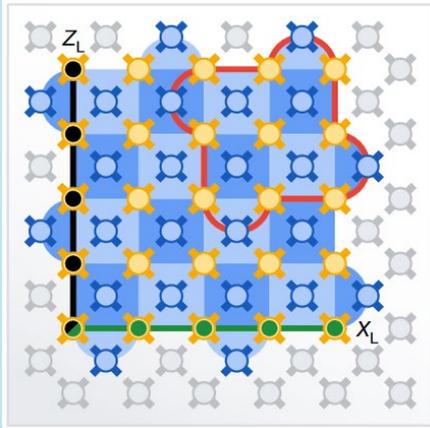

Error correction for quantum computing is challenging due to the quantum no-cloning theory, i.e. it is impossible to copy arbitrary quantum states. Stabilizer quantum error correction (QEC) codes circumvent this issue by encoding logical (fault-tolerant) qubit states in multiple entangled physical qubits, known as data qubits[18,22]. Ancillary physical qubits, referred to as measure qubits and placed alternated with data qubits, periodically measure the parity of data qubit ensembles. Changes in parity measurements reflect data qubit errors. The parity measurement results are analysed with a decoding algorithm, running in parallel on classical hardware, which determines the most probable physical error and calculates the appropriate compensation[29]. The surface code is one of the most investigated implementations for QEC due to its high tolerance of physical qubit error rates and limited qubit interconnection requirement of only nearest-neighbour interaction[29].

The surface code layout can be mapped to a 2D array of physical qubits, where every qubit interacts only with its four neighbouring qubits, as shown in the Box 1 inset figure. Data and measure qubits are shown in gold and blue, respectively. Light and dark blue tiles indicate the measure qubits used for Z and X gate error detection, respectively[21]. The total number of required physical qubits strongly depend on the physical qubit error rate, where a system with lower error rate requires fewer physical qubits[29]. Furthermore, the total number of required logical qubits requires on the quantum algorithm specifications and desired algorithm runtime. These intimate dependencies highlight the need for a holistic system design approach for fault-tolerant quantum computing.

Figure adapted from Ref. [21]

**Box 2**

# Interconnects: bandwidth, power consumption, and density

There are several interfaces important for fault-tolerant quantum computing, as shown in Fig. 1a with each interface requiring a physical connection to transport signals between adjacent layers. The differences in temperature, ranging from cryogenic to room temperature, bandwidth requirements, and space availability make it unlikely that a single interconnect technology will be sufficient for all connectivity needs throughout the full stack.

At the lowest level, interconnects between qubit devices and classical control electronics carry qubit control signals and qubit state readout signals. These interconnects couple to qubit devices operating at cryogenic temperatures between a few milli Kelvin and a few Kelvin. With qubit gate pitches ranging from ~10 nm to ~100 nm the wiring density will be highest at this layer. The high density and limited thermal budget of this interconnect make advanced technology nodes and 3D packaging techniques more likely candidates[194,195]. It's worth noting that challenges related to wiring density at this lowest level can be elevated with increased device uniformity, enabling parallel device control using a crossbar architecture similar to DRAM operation[41,61,62].

At the middle interface digital signals from the classical control electronics must be routed to the QEC encoder. The bandwidth of this interface will depend on the error correction algorithm clock cycle, which is bound by qubit coherence and practical computational time considerations on the lower side, and power consumption, available space, and qubit operation speed on the upper side. Importantly, advances in QEC can rapidly change requirements on the bandwidth, e.g. through optimized quantum low-density parity-check codes[196].

Finally, the top interface in the quantum computing stack connects the QEC encoder and quantum algorithm compiler. The signals are purely digital and the interface is most likely to be at room temperature.